\documentclass[onecolumn,10pt]{article}
\usepackage{amsmath}
\usepackage{amsfonts}
\usepackage{amssymb}
\usepackage{amsthm}
\usepackage{times}
\usepackage{cancel}
\usepackage{graphics}

\newcommand{\hil}[1]{\mbox{$\mathcal{#1}$}}

\newcommand{\ket}[1]{| #1 \rangle}

\usepackage{tikz}

\title{\textbf{Contextuality in Quantum Mechanics:\\ Testing the Klyachko Inequality}}
\author{Jeffrey Bub\thanks{email address: jbub@umd.edu} \\ \small \textit{Philosophy Department and Institute for Physical Science and Technology}\\  \small \textit{University of Maryland, College Park, MD 20742, USA}\\ Allen Stairs \thanks{email address: stairs@umd.edu}\\ \small \textit{Philosophy Department, University of Maryland, College Park, MD 20742, USA}  }
\date{}

\normalsize
\begin{document}

\maketitle

\begin{abstract}

\end{abstract} 
The Klyachko inequality is an inequality for the probabiities of the values of five observables of a spin-1 particle, which is satisfied  by any noncontextual assignment of values to this set of observables, but is violated by the probabilities defined by a certain quantum state. We describe an experiment between two entangled spin-1 particles to test contextuality via a related inequality. We point out that a test of contextuality by measurements on a single particle to confirm the Klyachko inequality requires an assumption  of non-disturbance by the measuring instrument, which is avoided in the two-particle experiment.

\bigskip


\section{The Klyachko Inequality}

In \cite{Klyachko2002,Klyachko2007}, Klyachko proposed a remarkably simple state-dependent version of the Kochen-Specker theorem. Recall that Kochen and Specker \cite{KochenSpecker} identified a finite  set of 1-dimensional projection operators on a 3-dimensional Hilbert space, in which an individual projection operator can belong to different orthogonal triples of projection operators representing different bases or measurement contexts, such that no assignment of 0 and 1 values to the projection operators is possible that is both (i) \emph{noncontextual} (i.e., each projection operator is assigned one and only one value, independent of context), and (ii) \emph{respects the orthogonality relations }(so that the assignment of the value 1 to a 1-dimensional  projection operator $P$ requires the assignment of 0 to any projection operator orthogonal to $P$). In a measurement of a  quantum system associated with a 3-dimensional Hilbert space, the system is required to produce a value for an observable represented by a 1-dimensional projection operator $P$ with respect to the measurement context defined by $P$ and its orthogonal complement $P^{\perp}$ in a nonmaximal measurement, or with respect to a measurement context defined by a particular orthogonal triple of projection operators in a maximal measurement. Unlike the situation in classical mechanics, different maximal measurement contexts for a quantum system are exclusive: they cannot all be embedded into one context. In this sense, measurement in quantum mechanics is contextual, and the distribution of measurement outcomes for a quantum state cannot be simulated by a noncontextual assignment of values to all observables, or even to certain finite sets of observables, by the Kochen-Specker theorem. 

Note that the contextuality of individual measurement outcomes is masked by the statistics, which is noncontextual:  the probability that a measurement of an observable corresponding to a projection operator $P$ yields the value 1 in a quantum state $\ket{\psi}$ is the same, irrespective of the measurement context, i.e., irrespective of what other projection operators are measured together with $P$ in the state $\ket{\psi}$. 
 
Klyachko derived an inequality for five observables, which is satisfied  by any noncontextual assignment of values to this set of observables, but is violated by the probabilities defined by a certain quantum state. Here is a sketch of Klyachko's derivation of the inequality.\footnote{This formulation of Klyachko's result, reproduced from  \cite{BubStairs2009},  owes much to a discussion with Ben Toner and differs from the analysis in \cite{Klyachko2002,Klyachko2007}.}

Consider a unit sphere and imagine a circle $\Sigma_{1}$ on the equator of the sphere with an inscribed pentagon and pentagram, with the vertices of the pentagram labelled in order 1, 2, 3, 4, 5 (see Fig. 1). 
 \begin{figure}[!ht]
    \begin{picture}(300,180)(-65,0)
\begin{tikzpicture}
\tikzstyle vertex=[circle,draw,fill=black,inner sep=1pt]
\path (0,0) coordinate (O);
\path (3*72+18:3cm) coordinate (P1);
\path (18:3cm) coordinate (P2);
\path (2*72+18:3cm) coordinate (P3);
\path (4*72+18:3cm) coordinate (P4);
\path (72+18:3cm) coordinate (P5);
\path (3*72+18:.4cm) coordinate (Q);
\node at (.25,.3)  {O};
\node at (-20:.3cm)  {$\theta$};
\path (3*72+18:3.5cm)
node {$1$};
\path (18:3.5cm) 
node {$2$};
\path (2*72+18:3.5cm) 
node {$3$};
\path (4*72+18:3.5cm)
node {$4$};
\path (72+18:3.5cm) 
node {$5$};
\draw[very thick] (P1) -- (P2) -- (P3) -- (P4) -- (P5) -- cycle;
\draw (P1) -- (P4) (P4) -- (P2)
(P2) -- (P5) (P5) -- (P3)
(P3) -- (P1);	
\draw (P1) -- (O) (O) -- (P2);
\draw (Q) arc (-125:10:.5cm);
\draw (0,0) circle (3cm);
\node[vertex] at (0,0) {};
\node[vertex] at (P1) {};
\node[vertex] at (P2) {};
\node[vertex] at (P3) {};
\node[vertex] at (P4) {};
\node[vertex] at (P5) {};
\end{tikzpicture}
    \label{fig1}
\end{picture}
\caption{Circle $\Sigma_{1}$ with inscribed pentagram}
 \end{figure}
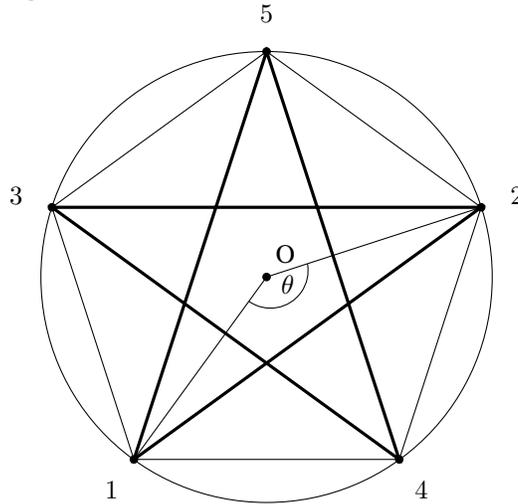

Note that the angle subtended at the center $O$ by adjacent vertices of the pentagram defining an edge (e.g., 1 and 2) is $\theta = 4\pi/5$, which is greater than $\pi/2$. It follows that if the radii linking O to the vertices are pulled upwards towards the north pole of the sphere, the circle with the inscribed pentagon and pentagram will move up on the sphere towards the north pole. Since $\theta = 0$ when the radii point to the north pole (and the circle vanishes), $\theta$ must  pass through $\pi/2$ before the radii point to the north pole, which means that it is possible to draw a circle $\Sigma_{2}$ with an inscribed pentagon and pentagram on the sphere at some point between the equator and the north pole, \emph{such that the angle subtended at $O$ by an edge of the pentagram is $\pi/2$}.  Label the centre of this circle $P$ (see Fig. 2; note that the line OP is orthogonal to the circle $\Sigma_{2}$ and is not in the plane of the pentagram). 

\begin{figure}[!ht]
    \begin{picture}(300,250)(-65,0)
\begin{tikzpicture}
\tikzstyle vertex=[circle,draw,fill=black,inner sep=1pt]
\path (0,0) coordinate (P);
\path (0,-5) coordinate (O);
\path (3*72+18:3cm) coordinate (P1);
\path (18:3cm) coordinate (P2);
\path (2*72+18:3cm) coordinate (P3);
\path (4*72+18:3cm) coordinate (P4);
\path (72+18:3cm) coordinate (P5);
\path (0,-4.2) coordinate (Q);
\node at (.25,.3)  {P};
\node at (.3,-4.8)  {O};
\node at (-.2,-4.4)  {$\phi$};
\path (3*72+18:3.5cm)
node {$1$};
\path (3*72+10:1.2cm)
node {$s$};
\path (18:3.5cm) 
node {$2$};
\path (25:1.5cm) 
node {$s$};
\path (-94:2cm) 
node {$r$};
\path (-40:.65cm) 
node {$\sqrt{2}$};
\path (2*72+18:3.5cm) 
node {$3$};
\path (4*72+18:3.5cm)
node {$4$};
\path (72+18:3.5cm) 
node {$5$};
\draw[very thick] (P1) -- (P2) -- (P3) -- (P4) -- (P5) -- cycle;
\draw (P1) -- (P4) (P4) -- (P2)
(P2) -- (P5) (P5) -- (P3)
(P3) -- (P1);	
\draw (O) -- (P) (O) -- (P1) (P) -- (P2) (P) -- (P1); 
\draw (Q) arc (90:125:.8cm);
\draw (0,0) circle (3cm);
\node[vertex] at (0,0) {};
\node[vertex] at (P1) {};
\node[vertex] at (P2) {};
\node[vertex] at (P3) {};
\node[vertex] at (P4) {};
\node[vertex] at (P5) {};
\node[vertex] at (O) {};

\end{tikzpicture}

\end{picture}
\caption{Circle $\Sigma_{2}$ with inscribed pentagram}
    \label{fig2}
\end{figure}
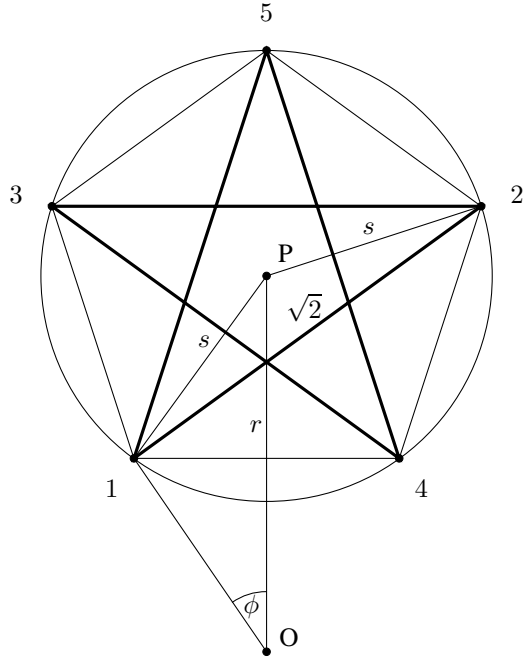

One can therefore define five orthogonal triples of vectors, i.e., five bases in a 3-dimensional Hilbert space $\hil{H}_{3}$, representing five different measurement contexts:
\[ \begin{array}{lll}
\ket{1},&\ket{2},&\ket{v} \\
\ket{2},& \ket{3}, & \ket{w}  \\
\ket{3}, & \ket{4}, & \ket{x} \\
\ket{4}, & \ket{5}, & \ket{y}  \\
\ket{5}, & \ket{1}, & \ket{z}
\end{array} \]
Here $\ket{v}$ is orthogonal to $\ket{1}$ and $\ket{2}$, etc. Note that each vector $\ket{1},\ket{2}, \ket{3}, \ket{4}, \ket{5}$ belongs to two different contexts. The vectors $\ket{u}, \ket{v}, \ket{x}, \ket{y}, \ket{z}$ play no role in the following analysis, and we can take a context as defined by an edge of the pentagram in the circle $\Sigma_{2}$, i.e., one of the edges 12, 23, 34, 45, 51.

Consider, now, assigning 0's and 1's to all the vertices of the pentagram in the circle $\Sigma_{2}$ noncontextually (i.e., each vertex is assigned a value independently of the edge to which it belongs), in such a way as to satisfy the orthogonality constraint that at most one 1 can be assigned to the two vertices of an edge (since, in a measurement of the observable corresponding an edge of the pentagram, the system is required to select one vertex of the edge, designated by the assignment of a 1 to the selected vertex and 0 to the remaining vertex, or to select no vertices, in which case the vector orthogonal to the edge is selected and assigned 1, and the two vertices of the edge are both assigned 0). 

It is  obvious by inspection that the orthogonality constraint can be satisfied noncontextually by assignments of zero 1's, one 1, or two 1's (but not by three 1's, four 1's, or five 1's). Call such assignments `charts' of type $C_{0}, C_{1}, C_{2}$, respectively. There are 11 such charts: one chart of type $C_{0}$ (corresponding to the assignment of 1 to the five vectors orthogonal to the five edges of the pentagram), five possible charts of type $C_{1}$  (each corresponding to the assignment of 1 to one of the five vertices of the pentagram, and 0 to the remaining vertices), and five possible charts of type $C_{2}$ (each corresponding to the assignment of two 1's to the vertices of one of the five edges of the \emph{pentagon} connecting the five vertices of the pentagram, i.e., the pentagon with edges 14, 42, 25, 53, 31 in the circle $\Sigma_{2}$).

If we label the possible charts with a hidden variable $\lambda_{j} \in \Lambda = \{\lambda_{j}: i = 1, \ldots, 11\}$, and average over $\Lambda$, then the probability that a vertex $v(i), 1 = 1, \ldots, 5$, is assigned the value 1 is given by:
\begin{equation}
p(v(i) =1) = \sum_{\Lambda} v(i|\lambda_{j})p(\lambda_{j})
\end{equation}
so:
\begin{eqnarray}
\sum_{i=1}^{5}p(v(i)=1) & = & \sum_{i=1}^{5}\sum_{\Lambda} v(i|\lambda_{j})p(\lambda_{j})\nonumber \\
& = & \sum_{\Lambda} (\sum_{i=1}^{5} v(i|\lambda_{j}))p(\lambda_{j}) \label{Klyachko}
\end{eqnarray}
Since, for a given $\lambda_{j}$, i.e., for a given chart,  $\sum_{i=1}^{5} v(i|\lambda_{j}) \leq 2$, and since 
$\sum_{j}p(\lambda_{j}) = 1$, it follows that 
\begin{equation}
\sum_{i=1}^{5}p(v(i)=1) \leq 2
\end{equation}

This is Klyachko's inequality: the sum of the probabilities assigned to the vertices of the pentagram  on the circle $\Sigma_{2}$, i.e., the probabilities that the projection operators corresponding to the vertices are assigned the value 1 in a measurement, must be less than or equal to 2, if the  value assignment  is noncontextual and satisfies the orthogonality constraint. Note that the  inequality follows without any assumption about the relative weighting of the charts, i.e., the probabilities $p(\lambda_{j})$, other than the assumption that these probabilities are independent of the measurement context. The relevance of this assumption will become clear in the following section.

Now consider a quantum system in the state defined by a unit vector that passes through the north pole of the sphere. This vector passes through the point $P$ in the center of the circle $\Sigma_{2}$. Call this state $\ket{\psi}$. A simple geometric argument shows that if probabilities are assigned to the 1-dimensional projectors defined by the vertices of the pentagram on $\Sigma_{2}$ by the  state $\ket{\psi}$, then the sum of the probabilities is greater than 2!

To see this, note that the probability assigned to a vertex, say the vertex 1, is:
\begin{equation}
|\langle 1|\psi\rangle|^{2} = \cos^{2} \phi
\end{equation}
where $\ket{1}$ is the unit vector defined by the radius from O to the vertex 1. Since the lines from the center $O$ of the sphere to the vertices of an edge of the pentagram on $\Sigma_{2}$ are radii of length 1 subtending a right angle, each edge of the pentagram has length $\sqrt{2}$. The angle subtended at $P$ by the lines joining $P$ to the two vertices of an edge is $4\pi/5$, so the length, $s$, of the line joining $P$ to a vertex of the pentagram is:
\begin{equation}
s = \frac{1}{\sqrt{2}\cos \frac{\pi}{10}}
\end{equation}
Now, $\cos \phi = r$, where $r$ is the length of the line $OP$, and $r^{2} + s^{2} = 1$, so:
\begin{equation}
\cos^{2} \phi = r^{2} = 1-s^{2} =  \frac{\cos \frac{\pi}{5}}{1+\cos\frac{\pi}{5}} = \frac{1}{\sqrt{5}}
\end{equation}
(because $\cos\pi/5 = \frac{1}{4}(1+\sqrt{5})$), and so:
\begin{equation}
\sum_{i=1}^{5} p(v(i) =1) = 5 \times \frac{1}{\sqrt{5}} = \sqrt{5} > 2
\end{equation}

\section{Testing the Inequality}
\label{sec:Test}

Consider testing the noncontextuality of quantum mechanics by preparing the state $\ket{\psi}$ as defined above and measuring the projection operators corresponding to the directions of the five vertices of the pentagram. Assuming the correctness of quantum mechanics, one will obtain the probabilities $p(v(i)) = \frac{1}{\sqrt{5}}$ so that $\sum_{i=1}^{5} p(v(i) =1)  > 2$, but such an inequality could also be obtained via noncontextual assignments of values to these projection operators, notwithstanding the argument in the previous section.

This would be possible if the probability distribution $p(\lambda_{j})$ were biased by the measurement. Suppose, for example, that measurements of the observables corresponding to the five edges of the pentagram are performed randomly, and 1's and 0's assigned to the vertices on the basis of these measurement outcomes. Suppose, also, that in a measurement of the observable corresponding to the edge 12, the measurement process biases the probabilities $p(\lambda_{j})$  in such a way that $p(\lambda_{j}) > 0$ only for charts assigning $v(1) = 1, v(2) = 0$ or $v(1) = 0, v(2) = 1$. To be specific, suppose the probability of charts assigning $v(1) = 1, v(2) = 0$ is 1/2 and the probability of charts assigning $v(1) = 0, v(2) = 1$ is 1/2, and all other charts are assigned probability 0. Suppose the analogous thing happens for measurements corresponding to the other edges of the pentagram as well. Then $p(v(i)=1) =1/2$ for all $i$, and so  $\sum_{i=1}^{5} p(v(i) =1)  =  \frac{5}{2}$, which is not only greater than 2, but even greater than the quantum mechanical value. Clearly, we can achieve the quantum mechanical value by assuming that in a measurement of the observable corresponding to the edge 12, the measurement biases the probabilities $p(\lambda_{j})$  in such a way that $p(\lambda_{j}) = \frac{1}{\sqrt{5}}$ for charts assigning $v(1) = 1, v(2) = 0$ or $v(1) = 0, v(2) = 1$, and $p(\lambda_{j}) = 1- \frac{1}{\sqrt{5}}$ for the remaining charts, and similarly for measurements corresponding to the other edges of the pentagram.

The possibility of such a bias or measurement disturbance is not inconsistent with any physical principle. Since the measurement interaction takes place at the site of the particle, one cannot exclude this possibility on the basis of a locality assumption, for example. So a test of contextuality via measurements on a single particle to confirm the Klyachko inequality would need to design an experiment to rule out or minimize this possibility. Here we propose an experiment involving two entangled spin-1 particles, $A$ and $B$, in which the correlations exclude a measurement bias. (We understand that a two-particle Klyachko experiment to test contextuality has been discussed independently by Anton Zeilinger and his group.\footnote{Private discussion with J. Bub during a visit to the Institute of Quantum Information and Quantum Optics, Austrian Academy of Sciences.})

Consider the maximally entangled state
\begin{equation}
\ket{\Psi} = \frac{1}{\sqrt{3}} \sum_{i = 1}^{3} \ket{\alpha_{i}}_{A}\ket{\alpha_{i}}_{B} \in \hil{H}_{3}\otimes\hil{H}_{3}  \label{eq:biorthog3}
\end{equation} 
where $\{\ket{\alpha_{1}}, \ket{\alpha_{2}}, \ket{\alpha_{3}}\}$ is an orthogonal basis in $\hil{H}_{3}$. A biorthogonal representation with equal coefficients takes the same form for any basis.  The marginal probabilities for the state $\ket{\Psi}$ for measurements corresponding to a vertex of the pentagram are all 1/3 for $A$ and also for $B$. Call this the `marginal constraint.' If a measurement of the projection operator corresponding to a particular vertex is performed on $A$, and a measurement of the projection operator corresponding to the same vertex is performed on $B$, then the two measurement outcomes always agree. Call this the `agreement constraint.' If a measurement corresponding to a particular vertex is performed on $A$, and a measurement corresponding to an orthogonal vertex is performed on $B$ (i.e., if the two measurements correspond to vertices that define an edge of the pentagram), then one never obtains the value 1 for both measurements. Call this the `orthogonality constraint.' Agreement and orthogonality are guaranteed by the correlations of the state $\ket{\Psi}$.

The proposed experiment is conducted in the following way: A choice of a measurement direction (one of the five vertices of the pentagram) is chosen randomly for $A$, and a corresponding measurement direction is chosen randomly for $B$. The results are recorded, and over many measurements a statistics is obtained for the marginal probabilities and for the following probabilities:
\begin{description}
\item[Case I:]
\begin{equation}
p(v(i)_{A}  = v(i)_{B}) 
\end{equation}
where $i = 1, \ldots, 5$ labels the same $A$-vertex and $B$-vertex.
\item[Case II:]
\begin{equation}
p(v(i)_{A}= 1, v(j)_{B}= 1) 
\end{equation}
where $i, j = 1, \ldots, 5$ label an $A$-vertex and a $B$-vertex that define an edge of the \emph{pentagram} (i.e., orthogonal directions).
\item[Case III:]
\begin{equation}
p(v(i)_{A}= 1, v(j)_{B}= 1) 
\end{equation}
where $i, j = 1, \ldots, 5$ label an $A$-vertex and a $B$-vertex that define an edge of the \emph{pentagon} (i.e., nonorthogonal directions). 
\end{description}

Case I tests the agreement constraint: these probabilities should all be 1 for the state $\ket{\Psi}$. Case II tests the orthogonality constraint: these probabilities should all be 0 for the state $\ket{\Psi}$. In Case III, the probabilities, for the state $\ket{\Psi}$, should all be equal to:
\begin{equation}
p = \frac{1}{3}(\frac{\sqrt{5} - 1}{2})^{2} \approx .12732 \label{eqn:p}
\end{equation}

To see this, note that the cosine of the angle, $\chi$, subtended at $O$ by two non-orthogonal states corresponding to two radii of the unit sphere subtending an edge of the \emph{pentagon} (see Fig. 3) is given by:\footnote{This is the inverse of the golden ratio, the limit of the ratio of successive terms in the Fibonacci series: $\tau = \frac{\sqrt{5} + 1}{2}$: $1/\tau = \tau -1$.}
\begin{equation}
\cos \chi =   \frac{\sqrt{5} - 1}{2}
\end{equation}
\begin{figure}[!ht]
    \begin{picture}(300,240)(-65,0)
\begin{tikzpicture}
\tikzstyle vertex=[circle,draw,fill=black,inner sep=1pt]
\path (0,0) coordinate (P);
\path (0,-5) coordinate (O);
\path (3*72+18:3cm) coordinate (P1);
\path (18:3cm) coordinate (P2);
\path (2*72+18:3cm) coordinate (P3);
\path (4*72+18:3cm) coordinate (P4);
\path (72+18:3cm) coordinate (P5);
\path (3*72+38:3.3cm) coordinate (Q);
\node at (.25,.3)  {P};
\node at (.3,-4.8)  {O};
\node at (-.9,-3.5)  {$\chi$};
\path (3*72+18:3.5cm)
node {$1$};
\path (3*72+10:1.2cm)
node {$s$};
\path (18:3.5cm) 
node {$2$};
\path (2*72+18:3.5cm) 
node{$3$};
\path (2*72+10:1.2cm) 
node{$s$};
\path (4*72+18:3.5cm)
node {$4$};
\path (72+18:3.5cm) 
node {$5$};
\draw[very thick] (P1) -- (P2) -- (P3) -- (P4) -- (P5) -- cycle;
\draw (P1) -- (P4) (P4) -- (P2)
(P2) -- (P5) (P5) -- (P3)
(P3) -- (P1);	
\draw (O) -- (P3);
\draw (O) -- (P1);
\draw (O) -- (P) (P) -- (P3) (P) -- (P1);
\draw (Q) arc (145:153:2cm);
\draw (0,0) circle (3cm);
\node[vertex] at (0,0) {};
\node[vertex] at (P1) {};
\node[vertex] at (P2) {};
\node[vertex] at (P3) {};
\node[vertex] at (P4) {};
\node[vertex] at (P5) {};
\node[vertex] at (O) {};

\end{tikzpicture}

\end{picture}
\caption{Pentagram  on $\Sigma_{2}$ showing angle $\chi$ between states $\ket{1}$ and $\ket{3}$}
    \label{fig3}
\end{figure}

From the geometry in Fig. 3:
\begin{equation}
\sin \frac{\chi}{2} = s \sin \frac{\pi}{5} = \frac{\sin \frac{\pi}{5}}{\sqrt{2}\cos \frac{\pi}{10}} = 
\sqrt{2} \sin\frac{\pi}{10} = \sqrt{2}\frac{\sqrt{5}-1}{4}
\end{equation}

Now consider simulating the marginal constraint, the agreement constraint, and the orthogonality constraint by noncontextual assignments of values to the vertices. To satisfy the agreement constraint, each pair of particles in the state $\ket{\Psi}$ would have to share the same chart (assuming the measurements are made within a time-interval that does not allow communication between the particles). To satisfy the orthogonality constraint, the charts would have to be of type $C_{0}, C_{1}$, or $C_{2}$. To satisfy the marginal constraint, a sequence of particle pairs  in the state $\ket{\Psi}$ on which measurements are to be performed would have to be associated with a shared mixture of these charts yielding the value 1/3 for the marginal probability of  obtaining 1 for a measurement direction corresponding to a vertex. That is, each particle pair in the sequence would have to share the same chart, and the charts shared by successive particle pairs in the sequence would have to range over charts of type $C_{2}, C_{1}$, and $C_{0}$, with the appropriate weights. 

For charts of type $C_{1}$, the probability of a vertex is 1/5, which is less than 1/3. So any mixture would have to include charts of type $C_{2}$. It is easy to see  that there is only one mixture, $M_{21}$, of charts of type $C_{2}$ and $C_{1}$, and one mixture, $M_{20}$, of charts of type $C_{2}$ and $C_{0}$ yielding a probability of 1/3 for each vertex:\footnote{The weights are the solutions to the linear equations $x\cdot\frac{2}{5} + y \cdot\frac{1}{5}  = \frac{1}{3}, x + y = 1$ for the mixture $M_{21}$, and $x\cdot\frac{2}{5} + y \cdot\ 0  = \frac{1}{3}, x + y = 1$ for the mixture $M_{20}$.}
\begin{enumerate}
\item[$M_{21}$:] 2/3 $C_{2}$, 1/3 $C_{1}$ 
\item[$M_{20}$:] 5/6 $C_{2}$, 1/6 $C_{0}$ 
\end{enumerate}
To obtain the probability 1/3, each particle  pair in the sequence is labelled with a shared hidden variable whose value selects a particular chart in the appropriate mixture, either $C_{2}$ or $C_{1}$ in the case of mixture $M_{21}$, or  $C_{2}$ or $C_{0}$ in the case of mixture $M_{20}$, where the values occur in the sequence with probabilities  corresponding to the mixture weights. Other mixtures yielding the marginal 1/3 involve appropriate mixtures of charts of all three types: $C_{2}, C_{1}$,and $C_{0}$.

For the mixture $M_{21}$, and vertices $i, k$ that label an $A$-vertex and a $B$-vertex, respectively, defining an edge of the \emph{pentagon}, we have:
\begin{eqnarray}
p(v(i)  =  1, v(k) = 1) & = & \frac{2}{3}\cdot \frac{1}{5} + \frac{1}{3}\cdot 0 \nonumber \\
& = & \frac{2}{15} \approx .133333 
\end{eqnarray}

For the mixture $M_{20}$, we have:
\begin{eqnarray}
p(v(i) =  1, v(k) = 1) & = &  \frac{5}{6}\cdot \frac{1}{5} + \frac{1}{6}\cdot 0 \nonumber \\
& = & \frac{5}{30} \approx .166666 
\end{eqnarray}

Since these probabilities are both greater than $p$ in (\ref{eqn:p}),  and since any mixture of charts $C_{2},C_{1}, C_{0}$ will yield probabilities between these values, it follows that the probabilities defined by noncontextual assignments of values to the vertices cannot simulate the probabilities defined by the quantum state $\ket{\Psi}$. 

Note that the agreement constraint forces the particles $A$ and $B$ to share the \emph{same} chart for each pair of measurements. A measurement bias for the selection of a particular chart is excluded here. The selection of $A$'s chart might be probabilistically biased by the measurement on $A$, but $A$'s chart has to be the same as $B$'s chart, and the selection of $B$'s chart could not be biased by $A$'s measurement if the measurements on $A$ and $B$ are space-like separated. Putting it somewhat anthropomorphically: the $B$-particle cannot `know' what measurement is performed on the $A$-particle, and so the selection of $B$'s chart cannot be biased by this measurement, and conversely for $A$. So this experiment, by contrast with the single particle experiment, is a crucial experiment for contextuality: it is impossible to satisfy the marginal constraint, the agreement constraint, and the orthogonality constraint by a noncontextual assignment of values to all the observables of the experiment, in such a way as to recover the probability $p$ defined by the state $\ket{\Psi}$.

One can put this in terms of an inequality. Denote the five edges of the \emph{pentagon} by $e_{i}, i = 1, \ldots, 5$, where the edge $e_{1}$ is defined by the vertex 1 for $A$ and 4 for $B$, the edge $e_{2}$ is defined by the vertex 4 for $A$ and 2 for $B$, etc. Let $p(e_{i})$ denote the probability of obtaining a 1 for both vertices of the $i$'th edge in a joint measurement on the two particles. Then we have:

For a noncontextual theory:
\begin{equation}
p(e_{1}) + p(e_{2})  + p(e_{3})  + p(e_{4})  + p(e_{5}) \geq \frac{2}{3}
\end{equation}

For the state $\ket{\Psi}$:
\begin{equation}
p(e_{1}) + p(e_{2})  + p(e_{3})  + p(e_{4})  + p(e_{5}) = \frac{5}{3}(\frac{\sqrt{5}-1}{2})^{2} \approx .636610 
\end{equation}

Note, also, that for any pair of measurements $X, X'$ corresponding to $A$-vertices, and any pair of measurements $Y, Y'$ corresponding to $B$-vertices, the Clauser-Horne-Shimony-Holt inequalities are satisfied. In the usual expression for the CHSH inequalities, the values of observables are $\pm 1$. Changing the units of the observables in the Klyachko experiment from 0, 1 to -1, +1, we have:
\begin{eqnarray}
\langle XY \rangle_{X=Y} & = & 1 \\
\langle XY \rangle_{\mbox{pentagram edge}} & = & -\frac{1}{3} \label{eqn:pentagram} \\
|\langle XY \rangle|_{\mbox{pentagon edge}} & \leq & \frac{1}{3}\label{eqn:pentagon} 
\end{eqnarray}
where $X, Y$ define a pentagram edge in (\ref{eqn:pentagram}), i.e., orthogonal directions, and $X,Y$ define a pentagon edge, i.e., nonorthogonal directions, in (\ref{eqn:pentagon}). So:
\begin{equation}
|\langle XY\rangle + \langle XY'\rangle +\langle X'Y\rangle - \langle X'Y'\rangle| \leq 2
\end{equation}
since at most two of these terms can be equal to 1, in which case the remaining two terms lie between $\pm 1/3$.

Since the correlations satisfy the CHSH inequalities but violate the Klyachko inequality, they test contextuality rather than nonlocality. That is, as far as the CHSH inequalities are concerned, the correlations are consistent with a local hidden variable theory for the observables measured in the experiment---they are inconsistent with a local hidden variable theory \emph{in which the value assignments are noncontextual}.

\section*{Acknowledgements}

Jeffrey Bub acknowledges support from  the University of Maryland Institute for Physical Science and Technology, and informative discussions on contextuality and the Klyachko inequality with Anton Zeilinger and Radek Rapkiewicz during a visit to the Institute for Quantum Optics and Quantum Information, Austrian Academy of Sciences, in May 2010.

\bibliographystyle{plain}
\bibliography{klyachko}

\end{document}